\documentstyle[epsfig,aps,preprint]{revtex} 
\def\laq{\raise 0.4ex\hbox{$<$}\kern -0.8em\lower 0.62 ex\hbox{$\sim$}}
\def\gaq{\raise 0.4ex\hbox{$>$}\kern -0.7em\lower 0.62 ex\hbox{$\sim$}}

\begin{document}
\bibliographystyle{unsrt}

\title{Primordial magnetic fields from inflation?}

\author{M. Giovannini\footnote{Electronic address: 
Massimo.Giovannini@ipt.unil.ch } and M. Shaposhnikov \footnote{Electronic 
address : Mikhail.Shaposhnikov@ipt.unil.ch }}

\address{{\it Institute of Theoretical Physics, University of Lausanne}}
\address{{\it BSP-1015 Dorigny, Lausanne, Switzerland}}

\maketitle

\begin{abstract}

The hot plasma above the electroweak scale contains (hyper) charged
scalar particles which are coupled to Abelian gauge fields. Scalars may
interact with gravity in a non-conformally invariant way and thus
their fluctuations can be amplified during inflation. These
fluctuations lead to creation of electric currents and produce
inhomogeneous distribution of charge density, resulting in the generation
of cosmological magnetic fields. We address the question whether
these fields can be coherent at large scales so that they may seed
the galactic magnetic fields.  Depending upon the mass of the charged
scalar and upon  various cosmological (critical fraction of energy
density in matter, Hubble  constant) and particle physics parameters
we found that the magnetic fields generated in this way are much
larger than vacuum fluctuations. However, their amplitude 
on cosmological distances is found to be too small for
seeding the galactic magnetic fields.

\end{abstract}
\vskip0.5pc
\centerline{Preprint Number: UNIL-IPT-00-09, April 2000 }
\vskip0.5pc
\noindent

\renewcommand{\theequation}{1.\arabic{equation}}
\setcounter{equation}{0}
\section{Formulation of the Problem} 

The idea that our galaxy could possess a magnetic field dates back 
to the (independent but simultaneous) works of Fermi \cite{fer}
(motivated by  the origin of high energy cosmic rays) and Schwinger
\cite{sch} (motivated by the origin of galactic synchrotron emission). Since
then, a lot of work  has been done both from the experimental and
theoretical side.  Listening to observations \cite{zel,kro} large
scale cosmological magnetic fields  can be estimated from Faraday
rotation effects and Zeeman splitting  of (hyperfine) spectral lines.
Listening to theory  the main puzzle is connected with the dynamical
origin  of large scale magnetic fields. 

It is usually assumed that the observed magnetic fields were
(exponentially) amplified by galactic dynamo mechanism from
pre-existing seed magnetic fields, coherent over scales of the
order of 100 pc at the time of galaxy formation. The amplitude of the
necessary seed fields is quite uncertain and depends on many details
of the dynamo mechanism \cite{zel,rus} and on cosmological parameters
\cite{turner,dav}. Typical vaules of the seeds
range from  $10^{-15}$ to $10^{-25}$ Gauss at the decoupling epoch.

In general, two types of ideas are considered for the explanation of
the seed magnetic fields. The first one is related to astrophysics. For 
instance, a Biermann battery mechanism can be postulated at the 
level of protogalaxy \cite{ost} and this will possibly lead 
to a mean current able to provide a source term 
for the evolution equations of the magnetic fields in the galactic 
plasma. The observation of large scale 
magnetic fields associated with clusters of galaxies \cite{tay} also 
suggested the appealing possibility of connecting (inter) galactic 
magnetic fields with active galactic nuclei \cite{col} (in this picture the 
dynamo mechanism is not essential since the rotation of a cluster 
is much smaller than the rotation of a galaxy). 
The second type of ideas relies on the 
 the interplay between particle physics and cosmology 
at different moments in the life of the Universe. In this 
framework various models were discussed such 
as  cosmological defects \cite{vil}, phase transitions \cite{PT} 
electroweak anomaly \cite{joy}, inflation \cite{turner,infl}, string cosmology
\cite{string}, temporary electric charge non-conservation \cite{dol1} or 
breaking of the gauge invariance \cite{turner}.

The main problem of most particle physics mechanisms of the origin of
seed fields is how to produce them coherently on {\em cosmological}
scales. All causal proposals (related, for example, to the bubble
collisions at the phase transitions) may produce sufficiently large
magnetic fields only on sub-horizon scales. Inflation, quite
effective in producing density fluctuations at super-horizon scales,
fails to amplify directly the vacuum fluctuations of the
electromagnetic field because of its conformally invariant coupling
to gravity.    

An attractive and very economical idea on the possible primordial
origin of the galactic magnetic fields was suggested
 in \cite{turner}. In
short, it is based on the following observation. While the coupling
of electromagnetic field to the metric and to the charged fields is
conformally invariant (this is not necessarily true in the models
with dynamical dilaton \cite{string}), the coupling of the charged 
scalar field to
gravity is not. Thus, vacuum fluctuations of the charged scalar field
can be amplified during inflation at super-horizon scales, leading
potentially to non-trivial correlations of the electric currents and
charges over cosmological distances. The  fluctuations of
electric currents, in their turn, may induce magnetic fields through
Maxwell equations at the corresponding scales. The role of the
charged scalar field may be played by the Higgs boson which couples to
the hypercharge field above the electroweak phase transition; the
generated hypercharged field is converted into ordinary magnetic
field at the temperatures of the order of electroweak scale. 

No detailed computations were carried out in \cite{turner} in order to
substantiate this idea. The suggestion of \cite{turner} was further
developed quite recently in \cite{cal} for the standard electroweak
theory with an optimistic conclusion that large scale  magnetic
fields can be indeed generated. In \cite{cal1} a supersymmetric model
was studied.

The aim of the present paper is to re-analyze this proposal and
compute the amplitude and the spectrum of seed magnetic fields that
may be generated because of amplification of zero-point fluctuations
of the scalar field during inflation. Our set-up resembles the one of
Ref. \cite{cal}.  We suppose that an inflationary phase was followed
by  a radiation dominated phase and we compute the charged particle
production associated with the change of the metric. This allows to
define the spectrum and magnitude of the current fluctuations at the
beginning of the radiation era. Taking the current distribution as an
initial condition, we study the plasma-physics problem of the 
relaxation of such an initial condition.  
We compute finally the magnetic fields, which
survived  possibly until the time of galaxy formation.

Depending upon the mass of the charged scalar and upon  various
cosmological (critical fraction of energy density in matter, Hubble 
constant) and particle physics parameters we found that the magnetic
fields generated in this way are much larger than vacuum
fluctuations, in agreement with qualitative conclusions of refs.
\cite{turner,cal,cal1}. However, in contrast with \cite{cal}, their
amplitude  on cosmological distances is found to be too small, by many 
orders of magnitude,  in order to seed the galactic
magnetic fields. We trace back this difference in the conclusions to the
discrepancy in the results obtained for the Bogoliubov 
coefficients (appearing in the problem of scalar particle
production) and to the treatment of the relaxation of electric currents in
conducting media during the radiation dominated epoch.

The plan of our paper is  the following. In Section II we will 
discuss the amplification of the charged scalar field in an 
expanding cosmological background. In Section III we will study the
connection between the  amplification of the charged scalar and the
production  of charge and current density fluctuations. In Section IV
we will develop a curved space description of the evolution of charge and
current  fluctuations within a  kinetic (Landau-Vlasov) approach. We
will apply our analysis to the physical  case of an
ultra-relativistic plasma prior to decoupling. Section V contains some
phenomenological  applications of our formalism. We will mainly be
concerned with  the generation of large scale magnetic fields in
various models  of cosmological evolution and with the possible
occurrence  of charge density and current fluctuations at large
scales.  Section VI contains  our concluding remarks.

\renewcommand{\theequation}{2.\arabic{equation}}
\setcounter{equation}{0}
\section{Amplification of a complex scalar field during inflation}
In this paper we will only consider scalar electrodynamics.
Possible generalizations of our results to theories containing more scalar
fields (for example, for electroweak theory or its extensions) are
straightforward. Since fermions are conformally coupled to gravity, 
their gravitational production is too small to generate any
substantial seed magnetic fields \cite{cal}.

Consider the action of a (massive) charged  scalar field minimally
coupled to the background geometry and to the electromagnetic field
(hypercharge field, if the standard model is assumed):
\begin{equation}
S = \int d^4 x \sqrt{-g} \bigl[ 
({\cal D}_{\mu} \phi)^{\ast} {\cal D}^{\mu} \phi - 
m^2 \phi^{\ast} \phi - 
\frac{1}{4} {\cal F}_{\alpha\beta}{\cal F}^{\alpha\beta} \bigr],
\label{actiong}
\end{equation}
where ${\cal D}_{\mu} = \partial_{\mu} - i e {\cal A}_{\mu}$,  
${\cal F}_{\mu \nu} = \partial_{[\mu} {\cal A}_{\nu]}$, 
$g_{\mu\nu}$ is the four-dimensional metric and $g$ its 
determinant.

We will suppose that the background geometry  is described, in conformal 
coordinates, by a Friedmann-Robertson-Walker (FRW) line element
\begin{equation}
g_{\mu\nu} = a^2 \eta_{\mu\nu},~~~ds^2= a^2(\eta) [d\eta^2 - d\vec{x}^2],
\label{metric}
\end{equation}
where $\eta_{\mu\nu}$ is the usual Minkowski metric.

It is convenient to introduce rescaled fields $\Phi= a \phi$ and
$A_\mu = a {\cal A_\mu}$. Correspondingly, we denote by $\vec{E}$ and
$\vec{B}$ the electric and magnetic fluctuations in curved  space.
They are related to the usual flat space fields $\vec{\cal E}$  and
$\vec{\cal B}$ by a time-dependent rescaling involving the scale
factor:
\begin{equation}
\vec{E} = a^2 \vec{{\cal E}}, ~~~\vec{B} = a^2 \vec{{\cal B}}.
\end{equation}
In terms of the rescaled fields the action is 
\begin{equation}
S = \int d^3 x d\eta\bigl[ 
\eta^{\mu\nu}D_{\mu} \Phi^{\ast} D_{\nu} \Phi + 
\bigl( \frac{a''}{a}- 
m^2 a^2\bigr)\Phi^{\ast} \Phi - \frac{1}{4} 
F_{\alpha\beta}F^{\alpha\beta}],
\label{action2}
\end{equation}
where the prime denotes differentiation  with respect to the
conformal time coordinate $\eta$ and  $D_{\mu} = \partial_{\mu} - i e
A_{\mu}$,  $F_{\mu \nu} = \partial_{[\mu} A_{\nu]}$. This is simply
the action of  electrodynamics in flat space-time with a time
dependent mass term for the scalar field. From this form of the action it is
obvious that there is no direct amplification of electromagnetic
fields during inflation. Moreover, for conformal coupling of the
scalar field to gravity the term proportional to $a''$ is absent and
the scalar particle production is supressed by the charged boson
mass. 

To compute the magnetic field fluctuations we will use a perturbative
approach. Namely, we will firstly compute the scalar particle production
omitting completely the coupling of the scalar field to the gauge
field (i.e. neglecting the back reaction, as it can be checked to be, 
a posteriori, self-consistent).  

The classical evolution equations for $\Phi$ in the case $A_\mu=0$ are 
simply given by:
\begin{equation}
\Phi''  - \nabla^2 \Phi - \frac{a''}{a} \Phi + m^2 a^2 \Phi=0.
\end{equation}
We will often use also decomposition of the complex scalar field in terms of
two real fields, $\Phi = (\Phi_1 +i\Phi_2)/\sqrt{2}$.

Once the background geometry is specified, 
the amplified inhomogeneities in the field $\Phi$ can be  computed.  
Suppose, as in
\cite{cal}, that the history of the Universe  consists of two
different epochs. A primordial phase,  whose background evolution is
not exactly known, and a  radiation dominated phase where the scale
factor $a(\eta)$ evolves  linearly in conformal time.  A continuous
(and differentiable)  choice of scale factors is then represented by
\begin{eqnarray}
&& a_{i}(\eta) = \biggl( - \frac{\eta}{\eta_1}\biggr)^{- \alpha}, 
~~~~\eta < - \eta_1,
\nonumber\\
&& a_{r}(\eta) = \frac{\alpha \eta + (\alpha + 1)\eta_1}{\eta_1},
~~~~\eta \geq - \eta_1, 
\label{a}
\end{eqnarray}
where $\alpha$ is some effective exponent parametrizing the  dynamics
of the primordial phase of the Universe.  Notice that if $\alpha = 1$
we have that the primordial  phase coincides with a de Sitter
inflationary epoch. In practice  we will consider $\alpha =1$ or
slightly deviating from it.

As a result of the change in the behaviour of the  scale factor
occurring at $-\eta_1$ the modes of $\Phi$ will be
parametrically amplified.  Defining $x_1 = k\eta_1$ and $\mu =
m\eta_1$ we can write the  Bogoliubov coefficients for $\alpha =1$
(standard inflation) and for cosmologically interesting case $x_1 \ll
1$ (long ranged fluctuations) and $\mu \ll 1$ (small scalar mass,
potentially giving rise to large scalar fluctuations):
\begin{eqnarray}
&&\alpha_{k} = e^{i \frac{\pi}{8}} \sqrt{\pi}\biggl\{ 
-\frac{ x_1^{ - \frac{3}{2}}}{ 2 \Gamma(\frac{3}{4}) \mu^{ \frac{1}{4}}} 
 + \frac{i\,x_1^{-\frac{1}{2}} }{2 \Gamma( \frac{3}{4}) \mu^{\frac{1}{4}}}
 + \biggl[ \frac{1}{2 \Gamma( \frac{3}{4}) \mu^{\frac{1}{4}}}
+ \frac{( i - 1) \mu^{1/4}}{\sqrt{2} \Gamma(\frac{1}{4})}\biggr]
 \sqrt{x_1}\biggr\}
+ {\cal O}(\mu^{\frac{5}{4}}),
\nonumber\\
&& \beta_{k} = e^{-i \frac{\pi}{8}} \sqrt{\pi}\biggl\{ 
\frac{x_1^{ - \frac{3}{2}} }{ 2 \Gamma(\frac{3}{4}) \mu^{\frac{1}{4}}} 
- \frac{i\,x_1^{-\frac{1}{2}} }{2 \Gamma( \frac{3}{4}) \mu^{\frac{ 1 }{4}}}
 + \biggl[ 
-\frac{1}{2 \Gamma( \frac{3}{4}) \mu^{\frac{1}{4}}}
+ \frac{( i + 1) \mu^{1/4}}{\sqrt{2} \Gamma(\frac{1}{4})}\biggr]
 \sqrt{x_1}\biggr\}
+ {\cal O}(\mu^{\frac{5}{4}}).
\label{bogds}
\end{eqnarray}
More general expressions can be found in Appendix A where  we also
give details about the calculation. The terms kept in the
expansion for small $x_1$ and $\mu$ is such that the unitarity condition
for the Bogoluibov coefficients is satisfied:  $|\alpha_{k}|^2 -
|\beta_{k}|^2 =1$. This property of the  truncation is quite
important: when discussing charge density fluctuations  we will see
that interesting cancellations  between the leading terms arise.

In the opposite limit,  for $k > 1/\eta_1$ the  mixing coefficients are
exponentially suppressed as $ \exp[ - \lambda k\eta_1]$ \cite{bir}. 
The coefficient $\lambda$ depends upon the details of the transition
between  the inflationary and radiation dominated phases.  The 
existence of an exponential suppression in the mixing coefficients
is  quite important from a general point of  view: it ensures  gentle
ultra-violet properties for the physical quantities  we ought to
compute.

We should mention that the leading contribution to the  amplification
coefficients has been computed in different  contexts \cite{TW}.
However, we are not only interested in the  leading behaviour of the
Bogoliubov coefficients but also  in the corrections whose
contribution is relevant  when the leading contribution cancels, as
in the case  of charge density fluctuations (see the following Section).

\renewcommand{\theequation}{3.\arabic{equation}}
\setcounter{equation}{0}
\section{Charge and current density fluctuations}
The Bogoliubov coefficients obtained in the previous Section specify
the probability of charged particle creation. The fluctuations in the
scalar field  induce also fluctuations in the electric current
associated with  the $U(1)$ symmetry of our action. The fluctuations 
in the charge and current density act as a source term for the evolution of the
fluctuations in the gauge fields. 

We will ignore, for the moment, the effect of the 
plasma conductivity which is
essential for the calculation of induced magnetic fields (it will be 
discussed in the following section)  and we will simply consider the structure
of the current-current correlators. In the curved space, the conservation
equation for the current is given by:
\begin{equation}
\frac{1}{\sqrt{- g}} \partial_{\mu} \bigl( \sqrt{- g} j^{\mu}\bigr) =0,
\end{equation}
where 
\begin{equation}
j^{\mu} = i e ( \phi^{\ast} \partial^{\mu} \phi - \phi \partial^{\mu}
 \phi^{\ast}).
\end{equation} 
It is convenient to introduce the rescaled current 
\begin{equation}
J^{\mu} = \sqrt{- g} g^{\mu\nu} j_{\nu}, 
\end{equation}
which can be expressed as 
\begin{equation}
J^{\mu} =
 e \biggl[\Phi_2 \partial^{\mu} \Phi_1 - \Phi_1 \partial^{\mu}
 \Phi_2\biggr]
\end{equation}
in terms of conformal fields. Notice that in this last expressions
the index appearing in the  derivatives is raised  (or lowered) using
the Minkowski metric and not the curved metric.

If we average $J^{\mu}$ on the vacuum we have clearly that  $\langle
J^{\mu} \rangle =0 $. However, the fluctuations  of the same quantity
for two space-time separated points are not zero.  By defining the
two-point functions of the field operators  $\Phi_1$ and $\Phi_{2}$
\begin{equation}
{\cal G}( x, y) = \langle \Phi_{1}(x) \Phi_{1}(y) \rangle = 
\langle \Phi_{2}(x) \Phi_2(y) \rangle,
\end{equation}
the two-point function of the charge and current density 
fluctuations can be written as 
\begin{equation}
\langle J_{\mu}(x) J_{\nu}(y) \rangle = 2e^2 \biggl\{ {\cal G}(x, y) 
\frac{\partial^2}{\partial x^{\mu} \partial y^{\nu}} {\cal G}(x, y) 
- \frac{\partial}{\partial y^{\nu}} {\cal G}(x,y) 
\frac{\partial}{\partial x^{\mu}} {\cal G}(x,y) 
\biggr\},
\end{equation}
where $x\equiv (\vec{x},\eta)$ and $y\equiv (\vec{y}, \tau)$ ($\eta$
and $\tau$  are two different conformal times). In the case of the
vacuum (no amplification took place) the two-point  functions are
simply
\begin{equation}
{\cal G}(x, y) = \frac{1}{2(2\pi)^3} \int \frac{d^3 k}{k_{0} } 
e^{i k(x- y)}.
\end{equation}
In the case of flat space-time the current density  fluctuations can
be expressed as 
\begin{equation}
\langle J_{\mu}(x) J_{\nu}(y) \rangle = \frac{e^2}{4 (2\pi)^6} \int
\frac{d^3 p}{p_0} \int \frac{d^3 k}{k_0} (p_{\mu} - k_{\nu} )^2 
e^{i (k+ p) (x-y)}.
\end{equation}
When the background passes from an inflationary phase to a radiation 
dominated phase we have instead that the two-point functions of  the
field operators can be written as 
\begin{eqnarray}
{\cal G}(x,y) &=& \int 
\frac{ d^3 k}{(2\pi)^3} \biggl[ |\alpha_{k}|^2 
g_{k}(\eta) g^{\ast}_{k} (\tau) 
+ |\beta_{k}|^2 
g^{\ast}_{k}(\eta) g_{k}(\tau) 
\nonumber\\
&+& \alpha_{k} \beta^{\ast}_{k}  
g_{k}(\eta) g_{k}(\tau) + \alpha^{\ast}_{k} \beta_{k} 
g^{\ast}_{k}(\eta)g^{\ast}_{k}( \tau)\biggr] 
e^{- i \vec{k} \cdot \vec{r}},
\label{corr}
\end{eqnarray}
where $g_{k}(\eta)$ are given by Eqs. (\ref{gfunc}) and where 
 $\vec{r} = \vec{x} - \vec{y}$ .

In order to define properly the charge and current density 
fluctuations at different scales it is helpful to introduce an averaging
procedure both over space  and over time. In the case of the charge
density fluctuations  we will define the charge fluctuations inside a
patch of volume $L^3$ and within a typical time ${\cal T}$ as  
\begin{equation}
Q^2_{L,{\cal T}}= \frac{1}{
{\cal T}^2}\int d^3 x \int d^3 y\int d\eta \int d\tau \langle 
J_{0}( \vec{x}, \eta) J_{0} (\vec{y}, \tau) 
\rangle W_{L,{\cal T}}(\vec{x},\eta- \xi_0)
W_{L,{\cal T}} (\vec{y},\tau -\xi_0),
\label{charge}
\end{equation}
where $W_{L,{\cal T}}$  are the smearing functions selecting the
contribution of the  correlator inside a given space-time region.  We
can choose the  smearing functions to be, for instance, the so-called
top hat function  which has a sharp edge. It is defined as  $  
W_{L,{\cal T}}(\vec{x},\eta)   =1$
for $|\vec{x}| \leq L $, $|\eta | < {\cal T}$ 
and it is equal to zero otherwise. This  choice is,
however, not so useful for our case. In fact, one can show that  the
Fourier transform of the top hat profile goes to zero, for large 
$k$, as $(k r_0)^{-3}$. Unfortunately this behaviour (ultimately
related to  the shrap edge of the profile) could  create spurious
effects in the charge and current density fluctuations.  A better
choice is, for our purposes, the gaussian smearing function 
\begin{equation}
W_{L,{\cal T}}(\vec{x}, \eta) = e^{ - \frac{|\vec{x}|^2}{2 L^2} - 
\frac{\eta^2}{2 {\cal T}^2}}.
\end{equation}
This expression smears the high frequency modes more  efficiently
than the top hat function. By inserting Eqs. (\ref{corr}) into Eq.
(\ref{charge}) we obtain, after regulatization over time, in the limit 
 $L \ll m^{-1}$ and for $\xi_0 \geq \eta_1$
\begin{equation}
Q^2_{L, {\cal T}} 
\simeq \frac{e^2}{ ( 2 \pi)^5} \int d^3 x \int d^3 y \int d^3 
k\int d^3 p | \alpha_{p} \beta_{k}  -
 \alpha_{k} \beta_{p}|^2 e^{ - \frac{ |\vec{x}|^2}{2 L^2} 
- \frac{|\vec{y}|^2}{2 L^2}} 
e^{ i (\vec{k} + \vec{p})\cdot (\vec{x} - \vec{y})}.
\label{chargecorr}
\end{equation}
A similar procedure can be carried on in the case of the  current
density fluctuation. 

As we will see in the next Section, the relevant quantity  to be
computed in order to estimate the size of the gauge field 
fluctuations is given by the trace of 
\begin{equation}
\langle (\vec{\nabla} \times \vec{J})_{k} (\vec{\nabla}\times \vec{J})_{l} 
\rangle = \epsilon_{i k a} \epsilon_{j l b} \frac{\partial^2}{\partial x^a 
\partial y^b} \langle J_{i}(\vec{x},\eta)J_{j}(\vec{y}, \tau) \rangle.
\end{equation}
Therefore, the regularized quantity we are interested in is
\begin{equation}
(\vec{\nabla}\times \vec{J})^2_{L,{\cal T}} = 
\frac{ e^2}{{\cal T}^2 }\int d^3 x \int d^3 y \int d \eta \int d \tau 
\langle (\vec{\nabla} \times \vec{J})_{k} (\vec{\nabla}\times \vec{J})_{k}
\rangle W_{L,{\cal T}}(\vec{x},\eta -\xi_0) W_{L,{\cal T}}(\vec{y},\tau - \xi_0). 
\label{curl}
\end{equation}
After regularization over time
we find, using Eq. (\ref{corr}) 
into Eq. (\ref{curl}) in the limit
$m\xi_0^2/\eta_1 \ll 1$, that for $L \gg m^{-1}$ and for 
$\xi_0 \geq \eta_1 $ our correlator becomes
\begin{equation}
(\vec{\nabla}\times \vec{J})^2_{L,{\cal T}} \simeq 
\frac{ e^2 }{(2\pi)^5} \frac{1}{m^2} \biggl( \frac{\eta_1}{\xi_0}\biggr)^2 
\int d^3 x\int d^3 y \int d^3 k\int d^3 p {\cal D}(k, p)
e^{ -\frac{|\vec{x}|^2}{ 2 L^2}-\frac{|\vec{y}|^2}{ 2 L^2} } 
e^{ i (\vec{k} + \vec{p})\cdot(\vec{x}- \vec{y})},
\label{curcor}
\end{equation}
where
\begin{equation}
{\cal D}(k,p) =[ k^2 p^2 - (\vec{k}\cdot 
\vec{p})^2] | \alpha_{k} + \beta_{k}|^2 |\alpha_{p} + \beta_{p} |^2. 
\end{equation}
\renewcommand{\theequation}{4.\arabic{equation}}
\setcounter{equation}{0}
\section{Magnetic fields from charge and current density fluactuations}
Once we know the charge and current fluctuations at the beginning of
the radiation dominated epoch we can compute the induced fluctuations
in the electromagnetic fields from the Maxwell equations. The
complete solution of this problem is hardly possible in a general
case because of the complicated (and model dependent) dynamics of the
inflaton decay and of the reheating of the Universe after inflation. We
shall use an extremely simplified picture of this process. Namely, we
will suppose that the plasma right after inflation is in thermal
equilibrium with some temperature $T$, up to small (at  super-horizon
scales) perturbations of the distribution functions leading to charge
and current fluctuations. Though this picture may be, in general,
 far from reality
 we expect it to reproduce the physics of the magnetic
field generation quite accurately, because the rate of equilibration
of the electromagnetic reaction is large compared with the rate of
the Universe expansion after inflation and may be bigger than the
rate of the inflaton decay. In fact, our result depends essentially
on the density of the charged particles and their typical momentum
only. Thus they may be qualitatively applicable even to situations with
large deviations from thermal equilibrium.

Within this set-up the problem can be related to the relaxation of
an initial perturbation in the distribution function. 
Consider an equilibrium homogeneous and
isotropic conducting plasma, characterized by a distribution function
$f_0(p)$ common for both positively and negatively charged
ultrarelativistic particles (for example, electrons and positrons) .
Suppose now that this plasma is slightly perturbed, so that the
distribution functions are
\begin{equation}
f_{+}(\vec{x}, \vec{p}, \eta) = f_0(p) + 
\delta f_{+}(\vec{x},\vec{p},\eta),
\,\,\,\,\,\,\,\,f_{-}(\vec{x},\vec{p}, \eta) =  
f_0(p) + \delta f_{-}(\vec{x},\vec{p},\eta),
\end{equation}
where $+$ refers to positrons and $-$ to electrons, and $\vec{p}$ is
the conformal momentum.  The Vlasov equation   defining the
curved-space evolution of the perturbed distributions can  be written
as  
\begin{eqnarray} 
&&\frac{\partial f_{+}}{\partial \eta} + 
\vec{v} \cdot \frac{\partial f_{+}}{\partial\vec{x}} + e ( \vec{E} + 
\vec{v}\times \vec{B}) \cdot\frac{\partial f_{+}}{\partial \vec{p}} = 
\biggl( 
\frac{\partial f_{+}}{\partial \eta}\biggr)_{\rm coll}
\label{Vl+},\\
&&\frac{\partial f_{-}}{\partial \eta} + 
\vec{v} \cdot \frac{\partial f_{-}}{\partial\vec{x}} - e ( \vec{E} + 
\vec{v}\times \vec{B})\cdot \frac{\partial f_{-}}{\partial \vec{p}} = 
\biggl(\frac{\partial f_{-}}{\partial\eta}\biggr)_{\rm coll}
\label{Vl-},
\end{eqnarray}
where the two terms appearing at the right hand side of each 
equation are the collision terms. This system of equation represents 
the curved space extension of the Vlasov-Landau approach  to plasma
fluctuations \cite{vla,lan}. All particle number densities here are
related to the comoving volume.

Notice that, in general  $\vec{v} = \vec{p}/\sqrt{\vec{p}^2 + m_{e}^2
a^2}$. In the  ultra-relativistic limit $\vec{v}= \vec{p}/|\vec{p}|$
and the  Vlasov equations are conformally invariant.  This  implies
that, provided we use conformal  coordinates and rescaled gauge
fields, the system of equations which we would have in flat space
\cite{lif} looks exactly the same  as the one we are discussing in a
curved FRW (spatially flat) background \cite{ber}.

The evolution equations of the gauge fields coupled to the two
Vlasov  equations can be written as 
\begin{eqnarray}
&& \vec{\nabla} \cdot \vec{E} = e \int d^3 p [f_{+}(\vec{x}, \vec{p},\eta) - 
f_{-}(\vec{x}, \vec{p},\eta)],
\nonumber\\
&& \vec{\nabla}\times \vec{E} + \vec{B}' =0,
\nonumber\\
&& \vec{\nabla}\cdot \vec{B} =0,
\nonumber\\
&& \vec{\nabla}\times \vec{B} -\vec{E}'= \int d^3 p \vec{v} 
[f_{+}(\vec{x}, \vec{p},\eta) - f_{-}(\vec{x}, \vec{p},\eta)].
\label{max}
\end{eqnarray}

Now, if $\delta f_{\pm}(\vec{x},\vec{p},\eta)$ at the beginning of the
radiation dominated epoch $\eta_0$ are known (their magnitude follows
from our computation of the Bogoluibov coefficients), and
$E(\vec{x},\eta_0)=B(\vec{x},\eta_0)=0$ initially, the magnetic
field at later times can be found from Eqs. (\ref{Vl+})--(\ref{max}).

This problem is solved in the Appendix B in a relaxation time
approximation for the  collision integral and under the 
assumption of small fluctuations in the distribution 
functions. Electric fields, as well as the charge
density perturbations quickly relax to zero during the time given by
the inverse conductivity of the plasma,  $\sigma \sim n_q/(\langle p
\rangle \nu)$, where $n_q \sim T^3$ is the density of the charged
particles and $\langle p \rangle \sim T$ is their average momentum, 
$\nu$ is  the frequency of collisions \footnote{All quantities here
are given for conformal time. For instance  the curved space
conductivity $\sigma$ is related to the  flat space one, $\sigma_c$ ,
through a rescaling which involves the  scale factor : $\sigma =
a\sigma_c$. Once we use our rescaled system $n,~T$ and  $\sigma$ 
do not change for adiabatic Universe expansion, provided that the effective 
number of massless degrees of freedom is constant. This is
different from Ref. \cite{cal} where ordinary (flat space) conductivity 
scales as temperature squared.}. As for the magnetic fields,  for
the most interesting case of large scales ($k^2 \ll
\sigma/\eta$  the result does not depend on  the frequency of
collisions $\nu$ and reads
\begin{equation}
\vec{B} \sim \frac{\langle p \rangle}{\alpha n_q}\vec{\nabla}\times
\vec{J},
\label{B2}
\end{equation}
where $\alpha$ is the fine structure constant.

\renewcommand{\theequation}{5.\arabic{equation}}
\setcounter{equation}{0}
\section{Estimates of magnetic field}

We are now ready to estimate the produced current  and  charge
density fluctuations. We will firstly  discuss the case of magnetic
fields assuming that  the primordial background is of de Sitter or
quasi-de Sitter type.  We will then move to estimate the charge
density fluctuations.

\subsection{De Sitter case} 
From Eq. (\ref{curcor}) we see that the correlation function  of
current density fluctuations is determined by the sum of the 
Bogoliubov coefficients. From Eq. (\ref{bogds}) we have
\begin{equation}
|\alpha_{k} + \beta_{k}|^2 |\alpha_{p} + \beta_{p}|^2  \sim
\biggl(\frac{ \pi^2}{  \,\Gamma(3/4)^4 } \sin^4{\frac{\pi}{8}}\biggr)
\mu^{- 1} |k \eta_1|^{-3} |p \eta_1|^{- 3},
\end{equation}
which implies
\begin{equation}
(\vec{\nabla}\times \vec{J})^2_{L,{\cal T}} =  \biggl( \frac{\eta_1}{\xi_0}\biggr)^2
\biggl(\frac{e^2}{(2
\pi)^5}  \frac{ \pi^2}{ \Gamma(3/4)^2} \sin^4{\frac{\pi}{8}}\biggr)
\frac{k_1^4}{\mu^3}  {\cal I}(L),
\end{equation}
where 
\begin{equation}
{\cal I}(L) = \int d^3 x \int d^3 y \int d^3 k \int d^3 p 
\frac{[k^2 \,p^2 -(\vec{k}\cdot\vec{p})^2]}{k^3 p^3} 
e^{ -\frac{|\vec{x}|^2}{2 L^2}} e^{- \frac{|\vec{y}|^2}{2 L^2}} 
e^{i (\vec{k} + \vec{p})\cdot (\vec{x} - \vec{y})}\simeq 0.9\times
(2\pi)^5 L^2.
\end{equation}

The final result for our correlator is 
\begin{equation}
\frac{(\vec{\nabla}\times \vec{J})^2_{L,{\cal T}}}{V^2} \simeq 
\biggl(0.9\,\frac{ e^2\,\pi^2}{ \Gamma(3/4)^4} \sin^4{\frac{\pi}{8}}\biggr)
\frac{k_1^4}{L^4} \mu^{-3},
\end{equation}
where $V \sim L^3$ is the typical volume of a region of size $L$; we take
$\xi_0 \sim \eta_1$ (larger values of $\xi_0$ give even smaller 
values of the magnetic fields). 

In order to get an estimate of the magnetic field, one should specify
the density of the charged particles and their average momentum, see
(\ref{B2}). A most realistic estimate would be to take $n \sim T^3$
and $\langle p \rangle \sim T$, where $T$ is the reheating
temperature.  Then we obtain for the gauge 
field fluctuations
\begin{equation}
\frac{B^2}{T^4} = 6.26 \, \biggl(\frac{m}{M_{P}}\biggr)^{-3} 
\biggl(\frac{H_1}{M_{P}}\biggr)^5 
\frac{1}{(L\,T)^4}.
\label{est1}
\end{equation}
In Eq. (\ref{est1})  $L$ is the coherence scale  of the magnetic
field. The ratio of the magnetic field to the $T^2$ is roughly 
constant during the Universe expansion  (if dynamo effects and the galaxy
collapse are discarded as well as annihilation of heavy particles) 
and may be taken at any time, provided the
coherence length $L$ is taken at the same epoch.  We will take 
$L\sim 1$ Mpc \cite{turner,cal} at the present
microwave background temperature 
\footnote{Equally, in agreement with the analysis in Ref.
\cite{dav}, one can take $L\sim 100$ pc at the galaxy formation time,
which gives essentially the same value for $LT$.}, 
which gives $LT \sim 3\times 10^{25}\,\,\,$. For a galactic
mass perturbation (of the order of $10^{12}$ solar masses, including
dark matter) the typical  length scale  is of the order of  $ 1.9
\times(\Omega_0 h^2)^{- 1/2}$ Mpc \cite{kt}. 
The gravitational  collapse of the
protogalaxies enhances the magnetic flux frozen in the primeval
galactic patch by roughly a factor of the order of  $10^3$. 

The obtained value for the magnetic field should be compared with the
values  of the magnetic field necessary to seed the galactic dynamo
mechanism. The differential  rotation of the galaxy introduces a
parity violating term in the MHD equations (the dynamo term). The
effect of this term exponentially amplify the  seed magnetic field by
a factor $e^{\Gamma t}$ \cite{zel,kro}. In this  amplification factor
enter two numbers: the galactic age (of the  order of $10$ Gyrs and
the dynamo amplification rate  ($\Gamma$) whose estimate is rather
uncertain: values  of the order $0.3$--$0.5$ Gyr for $\Gamma^{-1}$
are present  in the literature \cite{zel,rus}. Following a recent
analysis \cite{dav} we have that the required value for the seed
field can be expressed as 
\begin{eqnarray}
&& \frac{B_{\rm dec}}{T_{\rm dec}^2} \geq 5 \times 10^{-17},\,\,\,
{\rm for} \,\,\,\Gamma^{-1} = 0.5 \,{\rm Gyr}, 
\nonumber\\
&&  \frac{B_{\rm dec}}{T_{\rm dec}^2} \geq 2 \times 10^{-25},
\,\,\,  {\rm for}\,\,\, \Gamma^{-1} = 0.3\, {\rm Gyr}. 
\label{bounds}
\end{eqnarray}
where $T_{\rm dec}$ is the decoupling temperature.
These values have been obtained in the case of $\Omega_0 \sim 0.3$, 
$\Omega_{\Lambda} \sim 0.7$ and $h=0.65$ 
as a fiducial set of cosmological parameters \cite{tur,deb}. 
Notice that in the case 
of a flat Universe with $\Omega_0=1$ we would get values of 
$B_{\rm dec}/T_{\rm dec}^2$ close to $10^{-15}$. 
 
We want now to compare these values with the parameter space 
described by our estimate. If we take $m= 100$ GeV (the lower mass
limit for the Higgs boson) and if 
we assume  $H_1/M_{P}\simeq 10^{-6}$ (i.e. the maximal $H_1$
compatible with microwave background  anisotropies) we have that 
\begin{equation}
\frac{B_{\rm dec}}{T_{\rm dec}^2} \sim 5.77 \times 10^{-40},
\label{estim}
\end{equation}
twenty orders of magnitude smaller than the required value in order 
to seed the magnetic field of our galaxy.  One can argue that by
lowering the mass this estimate will  improve. This is not the case.
By taking $m \sim 1$  MeV (which is not realistic at all) we get 
$B_{\rm dec}/T_{\rm dec}^2 \sim 10^{- 32.5}$, still too small to be
relevant.  Notice, finally, that to tune $H_1$ does not help: since
it can only get  smaller that $10^{-6}$ it can only make the value of
the seed  even smaller than the values we just mentioned. 

The most conservative estimate of the magnetic field can be
obtained with the asumption that the number density of the charged
particles in the plasma is the number of gravitationally
created scalars,
\begin{equation}
n_{gr} = \int \frac{d^3 k}{(2\pi)^3} |\beta_k|^2 \sim 
\frac{1 }{8 \pi \Gamma^2(3/4)} 
\biggl(\frac{H_1}{m}\biggr)^{\frac{1}{2}}
H_1^3 \log{\biggl(\frac{H_1}{m}\biggr)}
\end{equation}
and their average momentum is simply $\langle p 
\rangle \sim H_1 \sim 100 {\rm GeV}$,
specific for the gravitational production.  If this is indeed the
case, the value of the magnetic field $B$ is larger by a factor 
\begin{equation}
8 \pi \Gamma^2(3/4)\biggl(\frac{m}{M_P}\biggr)^{1\over2}
\biggl(\frac{M_P}{H_1}\biggr)^{3\over2},
\label{ampl}
\end{equation}
if compared with the estimate (\ref{estim}). Following these considerations
the obtained magnetic field becomes
\begin{equation}
\frac{B}{T^2} \sim 10^2 \,\, \biggl(\frac{H_1}{m}\biggr) \frac{1}{(LT)^2}.
\label{est2}
\end{equation}
Even with the most optimistic numbers (i.e. $m\sim 100$ GeV and 
$ H_1 \sim 10^{-6} M_{P}$ ) we get that 
$B/T^2$ can be, at most, $10^{-38}$ over scales relevant 
for the dynamo action, with a minute gain, with respect to the 
estimate of Eq. (\ref{estim}). In priciple, one can think that 
even if the produced fields are too weak to 
turn-on the galactic dynamo they could be 
of some relevance for other processes occurring at different 
epochs in the life of the Universe. For instance, the 
electroweak horizon at the time of the electroweak 
phase transition \cite{gio,gio2} gives $L_{\rm ew}T_{\rm ew} 
\sim 10^{16}$. This would imply from Eq. (\ref{est2}) 
that $B_{\rm ew}/T_{\rm ew}^2 
\sim 10^{-19}$. 
At the electroweak epoch the smallest coherence lenght 
of magnetic fields is set by the diffusivity. Around 
the diffusivity scale $L_{\rm diff} T_{\rm ew} \sim
10^{8}$ \cite{gio}. Thus the obtained magnetic 
field can be, at  most, $B_{\rm ew} \sim 10^{-3} T_{\rm ew}^2$. 
 In order 
to have sizable effects on the phase diagram of the 
electroweak transition we should have, at least, that
$B_{\rm ew}/T_{\rm ew}^2 \geq 0.2$ \cite{kaj} 
so the produced fields are also 
too small in this context.

\subsection{Quasi de Sitter case}

Up to now we assumed that the primordial phase  in the evolution of
the Universe was of pure de Sitter type. In this subsection we are
going to study what happens is this requirement is relaxed. In
principle we know that during the inflationary  phase the scalar
field slightly decreases its value and, consequently,  the
inflationary curvature scale is not exactly constant but mildly 
decreasing as a function of time. In order  to account for the
decrease in the curvature and in the  scalar field we can define the
two slow-rolling parameters
\begin{equation}
\epsilon = - \frac{\dot{H}}{H^2} ,\,\,\, 
\lambda = \frac{\ddot{\phi}}{H \dot{\phi}},
\end{equation}
where $\phi$ denotes, in this Section, the inflaton and $H= \dot{a}/a$ 
is the Hubble 
parameter (the over-dot denotes differentiation with 
respect to the cosmic time). The important point for our purposes 
is that the slight decrease in the curvature corrects the evolution 
equation for the 
charged scalar. More specifically, we know that 
the dependence on the curvature appears in the mode function as $a''/a$. 
It is a simple exercise to show that 
\begin{equation}
\frac{a''}{a} = 2 a^2 H^2 ( 1 - \frac{\epsilon}{2}),
\end{equation}
which implies that the slow-rolling corrections 
will make the index appearing in the Hankel functions 
slightly larger
\begin{equation}
\rho = \frac{3}{2} + \epsilon.
\end{equation}
the case $\epsilon =0 $ corresponds to the pure de Sitter case. 
Clearly, from our general expressions of the Bogoliubov coefficients 
(see Eq.(\ref{bog1})) an increase in $\rho$ implies an increase 
of the Bogoliubov coefficients in the infra-red part of the spectrum. 

Therefore we want to estimate the magnetic fields in the case  where
the index $\nu$ is kept generic. For this  purpose, from Eq.
(\ref{bog1}) we have that 
\begin{equation}
|\alpha_{k} + \beta_{k}|^2 |\alpha_{p} + \beta_{p}|^2 \simeq
\biggl(
2^{4 \rho - 6}( \rho - \frac{1}{2})^4 \frac{\Gamma(\rho)^4}{\Gamma(3/4)^4}
 \sin^4{\frac{\pi}{8}} \biggr)\mu^{-1} 
| k\eta_1|^{- 2 \nu} |p\eta_1|^{-2\rho}.
\end{equation}
Following the same steps as in the pure de Sitter case we can estimate
that
\begin{equation}
\frac{B^2}{T^4} \simeq \biggl(
\frac{ 2^{ 4 \rho - 6} 
( \rho - \frac{1}{2})^4 \Gamma^4(\rho)}{ 2 \Gamma^4(3/4) e^2}
\sin^4{\frac{\pi}{8}} \biggr) \biggl(\frac{m}{M_P}\biggr)^{-3} 
\biggl(\frac{H_1}{M_{P}}\biggr)^{ 2 + 2 \rho}
(L \,T)^{4 \rho - 10}.
\label{nugeneric}
\end{equation}
If $\rho$ could be larger than $1.5$  the magnetic fields would also
be larger at the relevant scales.  

There are two relevant bounds on $\rho$. The first and obvious one  
comes from the energy density stored in the charged scalar  modes.
The energy density stored in the scalar field modes goes as  $m k^3
|\beta_{k}|^2$. This means, in the case of generic  $\rho$ that the
energy density (in critical units) is
\begin{equation}
\biggl(\frac{ m}{H_1} \biggr)^{\frac{1}{2}} 
\biggl( \frac{H_1}{M_{P}}\biggr)^{(\rho + \frac{1}{2}) }
\biggl(\frac{k}{T}\biggr)^{ 3 - 2 \rho}.
\end{equation}
If we take $m \sim 100 $ GeV and $\rho \sim 2$ we see that this
expression still gives a value $10^{-6}$ at the decoupling  scale.
Larger values of  $\rho$ could induce further  anisotropies in the
microwave background. So  we will assume $ 1.5 < \rho < 2$ which 
would be already enough to increase magnetic fields according to Eq. 
(\ref{nugeneric}).

The second class of bounds stems  from the fact that $\rho$ is
connected with the slow rolling parameters which are constrained. In
fact, the contribution to the  scalar spectral index  deduced from
the COBE data could be written,  in terms of the  slow-rolling
parameters, as \cite{kol,kol2}
\begin{equation}
n = 1 - 4 \epsilon + 2 \lambda.
\label{spectralindex}
\end{equation}
The values of $\epsilon$ and $\lambda$ are different  depending upon
the different models of background evolution, namely  upon the
different analytical forms of  the inflationary potential driving
inflation. This can be appreciated by  writing the equations of
motion of the inflaton in the  slow-rolling approximation
\begin{eqnarray}
&& 3 H \dot{\phi} + \frac{\partial V}{\partial\phi} \simeq 0,
\nonumber\\
&& M_{P}^2 H^2 \simeq V.
\end{eqnarray}
By using these two equations we can re-express
$\epsilon$ and $\lambda$ as 
\begin{eqnarray}
&& \epsilon = \frac{M_{P}^2}{6} \biggl(\frac{\partial 
\ln V}{\partial\phi}\biggr)^2,
\nonumber\\
&& \lambda = -\frac{M_{P}^2}{6} 
\biggl(\frac{\partial \ln V}{\partial\phi}\biggr)^2 + 
\frac{M_{P}^2}{3 V^2} \biggl(\frac{\partial^2  V}{\partial\phi^2}\biggr).
\end{eqnarray}
Clearly the values of $\epsilon$ and $\lambda$ depend upon the 
value of $\phi$. For instance, we could estimate the value of
$\epsilon$  coinciding with the value of the  field $50$ e-folds
before the end of inflation (corresponding to the moment  where the
large scales went out of the horizon) \cite{kol2}.  In this case, for
a power-law potential $V\sim \phi^{q}$  we have that 
\begin{equation}
\epsilon = \frac{q}{q + 200},\,\,\,\lambda= \frac{q - 2}{q + 200}.
\label{q}
\end{equation}
In the case of an exponential potential of the form  $V =\exp{[ 6
\phi^2/ (p M_{P}^2)]}$ we have that $\epsilon = \lambda = 1/p$.
Consequently,  from Eq. (\ref{spectralindex}), we have that $n =1 - (
2+ q)/100$ for power-law potentials, $ n = 1 - 2/p$ for exponential
potentials. The scale-invariant spectrum as it has been observed by
the DMR experiment  aboard the COBE satellite \cite{smo,teg} the
spectral index  can lie in the range $ 0.9 \leq n \leq 1.5$.  In
order to have more magnetic fields we should  increase $\rho$, namely
we should go for large $\epsilon$. The variation of  the spectral
constrains the maximal value of $\epsilon $. So if we take  $0.9$ as
the minimal value for $n$ we would have from Eq. (\ref{q}) that the
$q$  (for a power-law potential) is $q = 8$. But this would imply
that  the maximal $\epsilon$ is $0.03$ and our effective $\rho$ will 
be $1.53$. Too small to give relevant consequences in Eq.
(\ref{nugeneric}) for the magnetic fields generation. Similar
conclusions  could be reached in the case of power-law potentials. By
playing with the value of $\rho$ it is not possible  to enhance the
value obtained for large scale magnetic fields.

\subsection{Charge density fluctuations}

On the basis of the kinetic discussion the modes which survive  in
the plasma are the transverse ones. The  charge density fluctuations,
being associated with  the longitudinal modes will be dissipated
quite quickly in a typical  time of the order of the inverse
temperature. Still it is interesting to check if the charge density 
fluctuations are small. From our expression of the Bogoliubov
coefficients given in 
Eq. (\ref{bogds}) we have that 
\begin{equation}
| \alpha_{p} \beta_{k} - \alpha_{k}\beta_{p}|^2 = 
\frac{ \pi^2}{2 \Gamma(1/4)^2 \Gamma(3/4)^2}
\frac{|p\eta_1|}{|k \eta_1|^3}.
\end{equation}
We can insert this last expression into Eq. (\ref{chargecorr}) and perform
the 
integrations. The integrations over $x$ and $y$ are trivial. The 
integration over the moduli over the momenta leads to a  non trivial
integral which  can be exactly computed:
\begin{equation}
\int_0^{\infty} \frac{d z}{z} e^{ - z^2} \int_{0}^{\infty} w^2 
d w e^{- w^2} \sinh{ 2 w z} = -1/2.
\end{equation}
Defining then the electric charge density as 
\begin{equation}
n_e = \frac{Q_{L,T}}{L^6}
\end{equation}
we obtain that 
\begin{equation}
\frac{n_e}{n_{\gamma} } \sim 10^{-2} \biggl(\frac{H_1}{M_{P}}\biggr)^{1/2} 
(L T)^{-2},
\end{equation}
 where $n_{\gamma} \sim T^3$ is the photon density.
This value, for a length scale corresponding to the horizon at 
decoupling, would give $n_e/n_{\gamma} \sim 10^{-58}$. 
We want to stress that the  bounds \cite{ori,dol} on the 
electric charge fluctuations were derived by assuming 
that charge fluctuations would induce electric fields coherent over the 
whole horizon. These fields would cause some observable effect 
in the microwave background so that a constraint on the 
charge density could be derived. Again, on the basis of kinetic 
treatment of plasma fluctuations, we can say that electric fields 
dissipate as soon as the the plasma becomes conducting.
Therefore the effects on the microwave background are not present. 

\renewcommand{\theequation}{5.\arabic{equation}}
\setcounter{equation}{0}
\section{Conclusions}

In this paper we discussed the amplification and the fate of the
fluctuations of a charged scalar field in the inflationary Universe
scenario. These fluctuations may lead eventually to the  generation of
some magnetic fields in the Universe. We found that the produced
magnetic field are always much smaller 
than the most optimistic lower bounds required in order to seed the
galactic dynamo mechanism. Thus, the inflationary production of
charged scalars is unlikely to be responsible for the observed
galactic magnetic fields.
\section*{Acknowledgments}

We thank P. Tinyakov and C. Wagner for helpful discussions.

\newpage
\begin{appendix}

\renewcommand{\theequation}{A.\arabic{equation}}
\setcounter{equation}{0}
\section{Bogoliubov Coefficients} 
Defining $\Phi_1$ and $\Phi_2$ as the real and imaginary  part of
$\Phi$ as 
\begin{equation}
\Phi = \frac{1}{\sqrt{2}} ( \Phi_1 + i \Phi_2),
\end{equation} 
we assume that the background geometry evolves from $\eta\rightarrow
-\infty$  to $\eta\rightarrow + \infty$ for instance, according to
Eq. (\ref{a}).  In both limits we can define  a Fourier expansion of
$\Phi_1$ and $\Phi_2$ in terms of two  distinct orthonormal sets of
modes. By then promoting the classical  fields to quantum mechanical
operators in the Heisenberg  representation we can write, for $\eta
\rightarrow -\infty$
\begin{eqnarray}
&&
\Phi^{\rm in}_{1}(\vec{x},\eta) = 
\int \frac{d^3 k}{(2\pi)^{3/2}}\bigl[ a_{k} 
f_{k}(\eta) e^{i \vec{k}\cdot\vec{x}} + 
a_{-k}^{\dagger} f_{k}^{\ast}(\eta) e^{- 
i \vec{k}\cdot\vec{x}}\bigr],
\nonumber\\
&&
\Phi^{\rm in}_{2}(\vec{x},\eta) = 
\int \frac{d^3 p}{(2\pi)^{3/2}}\bigl[ b_{p} 
f_{p}(\eta) e^{i 
\vec{p}\cdot\vec{x}} + b_{-p}^{\dagger} f_{p}^{\ast}(\eta) 
e^{- i \vec{p}\cdot\vec{x}}\bigr].
\label{old}
\end{eqnarray}
where the two sets of creation and annihilation operators  (i.e. 
$[a_{k},a^{\dagger}_{-k}]$ and $[b_{p},b^{\dagger}_{-p}]$) are
mutually commuting. As $\eta \rightarrow + \infty$ $\Phi_1$ and
$\Phi_2$ can be expanded  in a second orthonormal set of modes 
\begin{eqnarray}
&&
\Phi^{\rm out}_{1}(\vec{x},\eta) = \int \frac{d^3 k}{(2\pi)^{3/2}}
\bigl[ \tilde{a}_{k} 
g_{k}(\eta) e^{i 
\vec{k}\cdot\vec{x}} + \tilde{a}_{-k}^{\dagger} g_{k}^{\ast}(\eta)
e^{- i \vec{k}\cdot\vec{x}}\bigr],
\nonumber\\
&&
\Phi^{\rm out}_{2}(\vec{x},\eta) = \int \frac{d^3 p}{(2\pi)^{3/2}}
\bigl[ \tilde{b}_{p} g_{p}(\eta) e^{i 
\vec{p}\cdot\vec{x}} + \tilde{b}_{-p}^{\dagger} g_{p}^{\ast}(\eta) 
e^{- i \vec{p}\cdot\vec{x}}\bigr].
\label{new}
\end{eqnarray}
Since both sets of modes are complete the old modes can be expressed
in  terms of the new ones 
\begin{equation}
f_{k}(\eta) = \alpha_{k} g_{k}(\eta) + \beta_{k} g_{k}^{\ast}(\eta).
\end{equation}
This transformation, once inserted back into Eq. (\ref{old}), implies
that
\begin{equation}
\tilde{a}_{k} = \alpha_{k} a_{k} + \beta_{k}^{\ast} a_{-k}^{\dagger}.
\label{def}
\end{equation} 
Notice that in order to preserve the scalar products in the old and
new  sets of orthonormal modes we have that the two complex numbers 
$\alpha_{k} $ and $\beta_{k}$ are subjected to the constraints 
$|\alpha_{k}|^2 - |\beta_{k}|^2 =1$. Exactly the same discussion
applies for  the field operator $\Phi_2$. Eq. (\ref{def}) is nothing
but the  well known Bogoliubov-Valatin transformation; $\alpha_{k}$
and $\beta_{k}$  are the Bogoliubov coefficients parametrizing the
mixing  between positive and negative frequency modes.

In order to ensure the continuity of the operators $\Phi_{1}$ and
$\Phi_{2}$  we have to match continuously the old mode functions with
the new ones. During the primordial phase of the Universe  the
evolution equations satisfied by the mode functions is 
\begin{equation}
\frac{d^2 f_{k}}{d\eta^2} + 
\biggl[ k^2 - \frac{\alpha(\alpha +1)}{\eta^2} 
+ \frac{\mu}{\eta_1^2}\bigl(\frac{\eta_{1}}{\eta}\bigr)^{2\alpha}\biggr] 
f_{k} =0,
\label{oldeq}
\end{equation}
where $\alpha = 1$ for a pure de Sitter  background $\mu = m/(H_1
a_1)= m \eta_1$. Notice that with $H$ we will  denote the Hubble
factor in cosmic time ( as usual the relation  between the Hubble
factor in conformal time, e.g. ${\cal H}$ and the Hubble  factor in
cosmic time is $H= {\cal H}/a$; during the de Sitter epoch ${\cal H}
\sim \eta^{-1}$). Notice that if the mass  of the charged scalar is
not of Planckian magnitude, $\mu \ll 1$. Moreover,  in the limit
where $\mu\geq 1$ one can argue that the amplified fluctuations  will
be exponentially suppressed. Since we want to explore  the situation
where the mass of the scalar field is of electroweak  order we will
always be (quite safely) in the limit $\mu \ll 1$. 
 
The exact solution of Eq. (\ref{oldeq}) which reduces, in the limit 
$\eta \rightarrow -\infty$, to the usual positive  frequency
Minkowski space solution is given by \cite{abr}
\begin{equation}
f_{k}(\eta)= \frac{1}{\sqrt{2 k}}\,\, p\,\,\, \sqrt{- x} \,\,\,
H^{(1)}_{\rho}(-x),  
\end{equation}
where $x = k \eta$ and $H^{(1)}_{\rho}$ is the first order Hankel
function.  Again, in the pure de Sitter case, $\rho = 3/2 \sqrt{ 1 -
(4/9) \mu^2}$. Since  $\mu $ is typically small, $\rho \simeq 3/2$ in
the pure de Sitter case. We denoted with $p$  a phase factor  which
we choose such that 
\begin{equation}
p = \sqrt{\frac{\pi}{2}}\,\,e^{i\frac{\pi}{4}( 1 + 2 \rho)}.
\end{equation}
With this choice of $p$ we have that  $f_{k}(\eta) \sim e^{- i
k\eta}/\sqrt{2 k}$ for $\eta \rightarrow - \infty$  \cite{abr}.

After the radiation dominated phase sets in (for $\eta > - \eta_1$ )
the evolution equation obeyed by the mode functions $g_{k}(\eta)$ is 
given by 
\begin{equation}
\frac{d^2 g_{k}}{d\eta^2} + \bigl[ k^2 
+ \frac{ \mu^2 (\eta + 2 \eta_1)^2}{\eta_1^4}\bigr] g_{k} =0.
\end{equation}
This last equation can be easily recast in the form of  a parabolic 
cylinder equation \cite{abr,erd}.  Defining $\gamma = \mu/\eta_1^2$
we can introduce two  new quantities, namely
\begin{equation}
z = \sqrt{2 \gamma} (\eta + 2 \eta_1),\,\,\,\, q = \frac{k^2}{2
\gamma}.
\label{neweq}
\end{equation}
Consequently, Eq. (\ref{neweq}) becomes
\begin{equation}
\frac{d^2 g_{k}}{d z^2} + \bigl[ q + \frac{z^2}{4} \bigr] g_{k} =0, 
\label{para}
\end{equation}
which is one of the canonical forms of the parabolic cylinder
equation  \cite{abr,erd}. The exact solutions which reduce to
positive and negative frequency modes for  $\eta \rightarrow +
\infty$ are 
\begin{eqnarray}
&& g_{k}(\eta) = \frac{ 1}{(2 \gamma )^{1/4} } \, e^{
i\frac{\pi}{8}}  D_{ - i q - \frac{1}{2}}( i e^{- i\frac{\pi}{4}} z),
\nonumber\\
&& g^{\ast}_{k}(\eta) = \frac{ 1}{(2 \gamma )^{1/4} } \, e^{
-i\frac{\pi}{8}}  D_{  i q - \frac{1}{2}}(  e^{- i\frac{\pi}{4}} z),
\label{gfunc}
\end{eqnarray}
where $D_{\sigma}$ are the parabolic cylinder functions  in the
Whittaker's notation \cite{erd}. Notice that with our choice  of
normalizations we have, in the limit $z \gg | q|$ and for $k^2 \eta_1 \ll m$, 
that \cite{abr,erd}
\begin{equation}
g_{k}(\eta) \sim \sqrt{\frac{ \eta_1}{2 m \eta}} 
e^{ - \frac{i}{2} \frac{m \eta^2}{\eta_1}}.
\label{g}
\end{equation}
In view of the actual calculation it is  worth  recalling  the exact
expressions of the parabolic cylinder functions  (in the Whittaker
form) in terms of confluent hypergeometric  (Kummer) functions $_1
F_1( a, b, x)$:
\begin{eqnarray}
D_{i q - \frac{1}{2}}( e^{- i \frac{\pi}{4}} z) &=&\sqrt{\pi} 
2^{\frac{i q}{2}-\frac{1}{4}}\,\,e^{\frac{i z^2}{4}} \biggl[
\frac{1}{ \Gamma(\frac{3}{4} - \frac{i q}{2})} {_1F_1}( \frac{1}{4} -
\frac{i q}{2},  \frac{1}{2}, - \frac{ i z^2}{2}) 
\nonumber\\
&&-  \frac{z ( 1 -
i)}{\Gamma(\frac{1}{4} - \frac{iq}{2})}  {_1F_1}( \frac{3}{4}
-\frac{i q}{2}, \frac{3}{2}, -\frac{i z^2}{2}) \biggr],
\nonumber\\
D_{- i q -\frac{1}{2} } ( i z e^{- i\frac{\pi}{4}}) &=&  \sqrt{\pi}
2^{ -\frac{i q}{2}  -\frac{1}{4}} e^{- i \frac{z^2}{4}}\biggl[
\frac{1}{\Gamma(\frac{3}{4} + \frac{i q}{2})} {_1F_1}(\frac{1}{4} +
\frac {i q}{2}, \frac{1}{2},  \frac{i z^2}{2}) 
\nonumber\\
&&- \frac{z ( 1 +
i)}{\Gamma(\frac{1}{4} + \frac{ i q}{2})} {_1F_1}( \frac{3}{4}
+\frac{i q}{2}, \frac{3}{2}, \frac{i z^2}{2})\biggr].
\end{eqnarray}

The Bogoliubov coefficients are obtained from
\begin{eqnarray}
&& f_{k}(-\eta_1) = \alpha_{k} g_{k}(-\eta_1) 
+ \beta_{k} g_{k}^{\ast}(-\eta_1), 
\nonumber\\
&& f'_{k}(-\eta_1) = \alpha_{k} g'_{k}(-\eta_1) 
+ \beta_{k} {g_{k}^{\ast}}'(-\eta_1),
\end{eqnarray}
which is a system of two equations in the two unknowns $\alpha_{k}$
and  $\beta_{k}$.

By solving this system we obtain an exact expression  for the
Bogoliubov coefficients which is, in general a function of two 
variables : $\mu= m\eta_1$ and $x_1 = k \eta_{1}$. Since $\mu \ll 1$
we can  expand the exact result in this limit and we obtain, in the
case of a  generic Bessel index $\rho$, 
\begin{eqnarray}
&&\alpha_{k} = \pi e^{ i \frac{\pi}{8}} \biggl\{  \frac{i}{ 
 \sqrt{2} \Gamma(\frac{3}{4})} S_2(x_1, \rho) \mu^{-\frac{1}{4}} + 
\frac{ (1 +i)}{2 \Gamma(\frac{1}{4})} [ S_1(x_1, \rho) + S_2(x_1, \rho)] 
\mu^{\frac{1}{4}} \biggr\} + {\cal O}(\mu^{\frac{5}{4}}),
\nonumber\\
&&\beta_{k} =  \pi e^{ -i \frac{\pi}{8}} \biggl\{ -\frac{ i}{ 
 \sqrt{2} \Gamma(\frac{3}{4})} S_2(x_1, \rho) \mu^{-\frac{1}{4}} + 
\frac{ (i -1)}{2 \Gamma(\frac{1}{4})} [ S_1(x_1, \rho) + S_2(x_1, \rho)] 
\mu^{\frac{1}{4}} \biggr\} + {\cal O}(\mu^{\frac{5}{4}}),
\label{bog1}
\end{eqnarray}
where $S_1(x_1, \rho)$ and $S_2(x_1,\rho) $ contain the explicit 
dependence upon the Hankel's functions:
\begin{eqnarray}
&& S_1(x_1, \rho) = e^{i \frac{\pi}{4} ( 1 + 2 \rho)}
H_{\rho}^{(1)}(x_1),
\nonumber\\
&& S_2(x_1, \rho) = \sqrt{x_1} e^{ i\frac{\pi}{4} ( 1 + 2 \rho)} 
\biggl[ \bigl( \rho + \frac{1}{2}\bigr) \frac{
H_{\rho}^{(1)}(x_1)}{\sqrt{x_1}} - \sqrt{x_1} H^{(1)}_{\rho + 1 } (x_1)
\biggr].
\label{s1s2}
\end{eqnarray}
If we are now specifically interested  in  the pure de Sitter phase
we can insert the value $\rho = 3/2$ into Eq. (\ref{s1s2}). Then, we
can  insert the obtained expressions into Eq. (\ref{bog1}) and we 
obtain the wanted Bogoliubov coefficients reported in Eqs.
(\ref{bogds}).

Notice that  it would not be correct to use the asymptotic solutions
(like the one reported  in Eq. (\ref{g})) in  order to compute the
Bogoliubov coefficients\footnote{We disagree with the calculation of
Ref. \cite{cal} where  the matching has been performed using
approximate mode functions. In our case, for small $\mu$ the leading
behaviour of the Bogoliubov coefficient goes as  $\mu^{-1/4}$. In the
case of \cite{cal} it goes as $\mu^{-1/2}$, an artifact  of the WKB
approximation.}. In fact Eq. (\ref{g}) can be viewed as the a
WKB-type solution of Eq.  (\ref{para}) which can be also written as 
\begin{equation}
\frac{d^2 g_{k}}{d\eta^2} + \omega_{k}^2(\eta) g_{k} =0,\,\,\,\,\,
\omega_{k}^2(\eta) = k^2 + m^2 a^2(\eta).
\label{wkbeq}
\end{equation}
By now postulating a WKB-type solution we have that
\begin{equation}
g_{k}(\eta) = \frac{1}{2 W(\eta)} e^{ - i \int^{\eta} W(\eta') d\eta'}.
\label{wkbsol}
\end{equation}
By now inserting the trial solution back to Eq. (\ref{wkbeq}) we get 
$W(\eta)$ is specified by the following non-linear relation
\begin{equation}
W^2(\eta) = \omega_{k}^2(\eta) - \frac{1}{2} \biggl[ \frac{ W''}{W} -
\frac{3}{2} 
\biggl(\frac{W'}{W}\biggr)^2\biggr].
\label{condition}
\end{equation}
This equation can be solved by iteration. If we keep the lowest order
we get  that 
\begin{equation}
W_{0}(\eta) \simeq \omega_{k}(\eta)
\end{equation}
and by using the explicit expression of the scale factor during  the
radiation dominated epoch we exactly get Eq. (\ref{g}). This
solution  is valid provided the corrections to the exact expression
of $W(\eta)$  are small, namely, provided, from eq, (\ref{condition})
\begin{equation}
\omega_{k}^2(\eta) \gg \frac{ 1}{2}\biggl[ \frac{ W_0''}{W_0} -
\frac{3}{2}  \biggl(\frac{W_0'}{W_0}\biggr)^2\biggr].
\end{equation}
This last inequality, using the explicit expression of
$\omega_{k}(\eta)$,  implies that
\begin{equation}
k^2 \eta^2 + m^2 \eta^2 \biggl( \frac{\eta+ 2 \eta_1}{\eta_1}\biggr) \gg 1.
\end{equation}
Now we can see that this inequality is clearly satisfied for $\eta 
\rightarrow + \infty$. However, for $\eta \sim \eta_1$, this
inequality  would imply $m \eta_1 > 1$ (since $k \eta_1 <1$).
Therefore, in order  to be consistent with the requirement that $m
\eta_1 <1 $ we have to use  the WKB-type solution only for large
(positive) $\eta$.

\renewcommand{\theequation}{B.\arabic{equation}}
\setcounter{equation}{0}
\section{Vlasov-Landau approach to electromagnetic field fluctuations.} 
The purpose of this appendix is to give details  concerning the
derivation of the relation between  the electromagnetic field
fluctuations and the  initial fluctuations in the current (or charge)
density profile.

By subtracting Eqs. (\ref{Vl+}) and (\ref{Vl-})  we obtain the
equations relating the fluctuations of  the distributions functions
of the charged particles present in the plasma  to the induced gauge
field fluctuations:
\begin{eqnarray}
&&\frac{\partial}{\partial\eta} f(\vec{x}, \vec{p},t) + 
\vec{v}\cdot \frac{\partial }{\partial\vec{x}} f(\vec{x},\vec{p},t) 
+ 2 e \vec{E}\cdot 
\frac{\partial f_0}{\partial \vec{p}} =- \nu(p) f,
\nonumber\\
&& \vec{\nabla} \cdot \vec{E} = e \int d^3 p f(\vec{x},\vec{p},\eta),
\nonumber\\
&& \vec{\nabla}\times \vec{E} + \vec{B}' =0,
\nonumber\\
&& \vec{\nabla}\cdot \vec{B} =0,
\nonumber\\
&& \vec{\nabla}\times \vec{B} -\vec{E}'= \int d^3 p \vec{v} 
f(\vec{x}, \vec{p},\eta),
\label{Vlasov}
\end{eqnarray}
where $f(\vec{x}, \vec{p}, \eta) = \delta f_{+}(\vec{x},\vec{p},\eta)
-  \delta f_{-}(\vec{x},\vec{p},\eta)$ and $\nu(p)$ is a typical
frequency of collisions. 

We can solve this system \cite{lif} by taking the Fourier transform
of the  space-dependent quantities and the Laplace transform of the 
time-dependent quantities:
\begin{eqnarray}
&&\vec{E}_{\vec{k} \omega} = \int_{0}^{\infty} d\eta e^{i \omega \eta} 
\int d^3 x e^{ - i \vec{k}\cdot \vec{x}} \vec{E}(\vec{x},\eta), 
\nonumber\\
&& \vec{B}_{\vec{k} \omega} = \int_{0}^{\infty} d\eta e^{i \omega \eta} 
\int d^3 x e^{ - i \vec{k}\cdot \vec{x}} \vec{B}(\vec{x},\eta), 
\nonumber\\
&& f_{\vec{k} \omega} 
 = \int_{0}^{\infty} d\eta e^{i \omega \eta} 
\int d^3 x e^{ - i \vec{k}\cdot \vec{x}} f(\vec{x},\vec{p},\eta).
\end{eqnarray}
We have now to specify, at the initial  time, the  form of the
perturbed distribution function which can be derived from  the
amplification studied in the previous Section.  We will call 
$g_{\vec{k}}(\vec{p})$ the initial profile of the distribution
function. Eq. (\ref{Vlasov}) can then be re-written as 
\begin{eqnarray}
&& - g_{\vec{k}}(\vec{p}) + i (\vec{k}\cdot \vec{v} - \omega) 
f_{\vec{k}\omega}(\vec{p}) + 2 e \vec{E}_{\vec{k} \omega} \cdot 
\frac{\partial f_0}{\partial\vec{p}} = -\nu f,
\label{vl1}\\
&& i \vec{k} \cdot \vec{E}_{\vec{k} \omega} = e 
\int f_{\vec{k} \omega}(\vec{p}) d^3 p,
\label{vl2}\\
&& i \vec{k} \cdot \vec{B}_{\vec{k} \omega} =0,
\label{vl3}\\
&& \vec{B}_{\vec{k} \omega} = \frac{1}{\omega} \vec{k}\times 
\vec{E}_{\vec{k} \omega},
\label{vl4}\\
&&i \omega \bigl( 1 - \frac{k^2}{\omega^2}\bigr) \vec{E}_{\vec{k}
\omega}   + \frac{i}{\omega} \vec{k} (\vec{k}\cdot \vec{E}_{\vec{k}
\omega})  = \int d^3 p ~\vec{v} ~f_{\vec{k} \omega}(\vec{p}),
\label{vl5}
\end{eqnarray}
where eq. (\ref{vl5}) has been obtained by using Eq. (\ref{vl4}) in
the  (Fourier  and Laplace) transformed of the last of Eqs.
(\ref{Vlasov}). The Gauss constraint at $\eta=0$ implies that 
\begin{equation}
i \vec{k} \cdot \vec{E}_{0}(\vec{k}) = e \int d^3 p
~g_{\vec{k}}(\vec{p}).
\label{gauss}
\end{equation}
If we start, at the initial time, with a given profile of
fluctuations fluctuation the Gauss constraint determines the initial
value of the  electric field. The magnetic field fluctuations  are
consistently equal to zero. 

We can now separate the  electric field in its polarizations parallel
and transverse to the  direction of propagation of the fluctuation. 
The transverse current provides a source for the evolution 
of transverse electric field fluctuations 
\begin{equation}
i \omega \bigl( 1 - \frac{k^2}{\omega^2} \bigr)  \vec{E}_{\vec{k}
\omega}^{\perp} = e \int d^3 p f_{\vec{k} \omega}(\vec{p})
\vec{v}_{\perp},
\label{perp}
\end{equation}
whereas the charge fluctuations provide a source for 
 the evolution of longitudinal  electric field fluctuations
\begin{equation}
i \vec{k} \cdot \vec{E}_{\vec{k} \omega}^{\parallel} = e \int d^3 p 
f_{\vec{k} \omega} (\vec{p}). 
\label{parallel}
\end{equation}
In Eqs. (\ref{perp}) and (\ref{parallel}) we defined the 
longitudinal part of the electric field fluctuations and the 
transverse electric field as
\begin{equation}
\vec{E}_{\vec{k} \omega}^{\perp} = \vec{E}_{\vec{k} \omega}^{\perp}
-  \frac{\vec{k}}{|\vec{k}|^2}  (\vec{k} \cdot \vec{E}_{\vec{k}
\omega}),\,\,\,\,\,\, \vec{E}_{\vec{k} \omega}^{\parallel} =
\frac{\vec{k}}{|\vec{k}|^2}  (\vec{E}_{\vec{k} \omega} \cdot
\vec{k}).
\end{equation}
The solution of  Eq. (\ref{vl1}) is given by  
\begin{equation}
f_{\vec{k} \omega}(\vec{p}) =  \frac{1}{i( \vec{k} \cdot \vec{v} -
\omega -i\nu)} \bigl[ g_{\vec{k}}(\vec{p}) - 2 e  \vec{v}\cdot
\vec{E}_{\vec{k} \omega} \frac{\partial f_0}{\partial p} \bigr],
\label{f}
\end{equation}
where we used that $\partial f_0/\partial \vec{p} \equiv 
\vec{v} \partial f_0/\partial p$. 
The longitudinal  and transverse components of the electric
fluctuations can be obtained  by inserting Eq. (\ref{f}) into Eqs.
(\ref{perp})-(\ref{parallel})
\begin{eqnarray}
&& |\vec{E}_{\vec{k} \omega}^{\parallel}| = \frac{e}{ 
i k \,\,\, \epsilon_{\parallel} } \int d^3 p 
\frac{g_{\vec{k}}(\vec{p})}{i (\vec{k}\cdot \vec{v} - \omega -i\nu)},
\label{par2}\\
&& \vec{E}_{\vec{k} \omega}^{\perp} = 
\frac{e \,\,\,\omega}{\omega^2 \epsilon_{\perp} - k^2} 
\int d^3 p \vec{v}^{\perp} 
 \frac{ g_{\vec{k}}(\vec{p})}{(\vec{k}\cdot \vec{v} - \omega-i\nu)}, 
\label{perp2}
\end{eqnarray}
where $\epsilon_{\parallel}$ and $\epsilon_{\perp}$ are,
respectively,  the longitudinal and transverse part of the
polarization tensor
\begin{eqnarray}
&&\epsilon_{\parallel}(k, \omega) = 1 - \frac{ 2 e^2 }{k^2} \int d^3 p 
\frac{\vec{k}\cdot \vec{v}}{ (\vec{\vec{k}} \cdot \vec{v} - \omega - i \nu)}
 \frac{\partial f_0}{\partial p},
\label{polpar}\\
&& \epsilon_{\perp} (k, \omega) = 1 - \frac{e^2}{ \omega} \int d^3 p 
\frac{ \vec{v}_{\perp}^2 }{(\vec{k}\cdot \vec{v} - \omega - i \nu)} 
\frac{\partial 
f_0}{\partial p}.
\label{polperp}
\end{eqnarray}
Now, the general expression for the generated magnetic field is 
\begin{equation}
\vec{B}_{\vec{k} \omega} = \frac{ e }{ \omega^2 
\epsilon_{\perp}(k,\omega) - k^2} 
\int d^3 p [\vec{v} \times \vec{k}] \frac{g_{\vec{k}}(\vec{p}) }{(\vec{k} 
\cdot \vec{v} - \omega -i\nu)}.
\label{Bg}
\end{equation}

The space-time evolution of the magnetic fluctuations can be
determined by performing the inverse Laplace and  Fourier transforms:
\begin{equation}
\vec{B}(\vec{x},\eta) = 
\int e^{- i \omega\eta}\frac{e~d\omega }{\omega^2 
\epsilon_{\perp}(k,\omega) - k^2 } \int d^3 k 
e^{ i \vec{k}\cdot \vec{x} } [\vec{v}\times \vec{k}]\int d^3 p 
\frac{ g_{\vec{k}}(\vec{p})}{ (\vec{v}\cdot \vec{k} - \omega -i\nu)}.
\label{bf}
\end{equation}

In order to perform this integral, the explicit relations for the
polarization tensors should be given. They depend on the equilibrium
distribution function $f_0(p)$, 
which we take to be\footnote{ 
Notice that most of  our considerations  can be easily
extended to the case of a Bose-Einstein or Fermi-Dirac  distribution.
What is important, in our context, is the analytical  structure of
the polarization tensors and this is the same  for different
distributions \cite{fra}.} 
\begin{equation}
f_{0}(p) = \frac{n_{q}}{8 \pi T^3} e^{-\frac{p}{T}},
\label{dist0}
\end{equation}
where $T$ is the equilibrium temperature, $p$ is the modulus of the 
momentum and $n$ is the equilibrium (thermal) density of charged
particles in the plasma. The normalization is chosen in such a way 
that $\int d^3 p f_0(p)= n_{q}$.

Then we have for transverse polarization
\begin{equation}
\epsilon_{\perp}(k, \omega) = 1 +   \frac{e^2 \,\,
n_q}{2 \omega k T} \biggl\{  \bigl[ 1 - \frac{(\omega+i\nu)^2}{k^2} \bigr]
\ln{\frac{k - \omega-i\nu}{k + \omega + i\nu}} - 2 \frac{\omega+i\nu}{k}\biggr\},
\label{perpapp}
\end{equation}
and for the longitudinal polarization:
\begin{equation}
\epsilon_{\parallel}(k, \omega) = 1 +  
\frac{e^2 \,\, n_q}{ k^2 T} \biggl\{ 
 2 +\frac{\omega+i\nu}{k} \ln{\frac{k - \omega-i\nu}{k + \omega +i\nu}}
\biggr\}.
\label{parallapp}
\end{equation}

Consider now the case of very small momenta $k\ll \omega$ and $\omega
\ll \nu$, relevant for long-ranged magnetic fields. Then the
computation of the integral (\ref{bf}) in the large time limit and with
the use of explicit form of the transverse polarization tensor in
(\ref{perpapp}) gives \footnote{For small $k$,  the equation  $\omega^2
\epsilon_{\perp}(k,\omega) - k^2=0$ defining the poles of the 
inverse Laplace transform implies $\omega \sim i k^2/\sigma$.}: 
\begin{equation}
B(\vec{x},\eta) \simeq \frac{T}{4\pi\alpha n_q} \exp{(-k^2 \eta/\sigma)}
\vec{\nabla} \times \vec{J},
\end{equation}
where $\sigma$ is the plasma conductivity in the relaxation time
approximation,
\begin{equation}
\sigma = \frac{2 e^2 n_q}{\nu T} 
\end{equation}
and initial electric current is given by
\begin{equation}
\vec{J}(\vec{x}) = \int d^3 p \,\vec{v}\, g_{\vec{k}}(\vec{p}) ~.
\end{equation}
 
In closing our discussion of the Vlasov equation  we want to briefly
comment about the validity of our approach. The obtained  results
assumed that the linearization of the Vlasov equation is consistent 
with the physical assumptions of our problem. This is indeed the
case.  In order to safely linearize the Vlasov equation we have to 
make sure that the perturbed distribution function of the charge 
fluctuations is always smaller than the first order of the 
perturbative expansion (given by the distribution of Eq.
\ref{dist0}).  In other words we have to make sure that
\begin{equation}
|\delta f_{+}(\vec{x},\vec{p},\eta)| \ll f_{0}(\vec{p}),\,\,\,
|\delta f_{-}(\vec{x},\vec{p},\eta)| \ll f_{0}(\vec{p}).
\end{equation}
These conditions imply that
\begin{equation}
\frac{e \vec{E}_{\vec{k} \omega}}{|\vec{k}\cdot \vec{v} - \omega-i\nu|}\cdot
\frac{\partial f_0(\vec{p})}{\partial \vec{p}} \ll f_0(\vec{p}).
\label{cond}
\end{equation}
If we now define the relativistic plasma frequency as
\begin{equation}
\omega^2_{p} = \frac{2 \,e^2\, n_q}{3 \,T}, 
\end{equation}
we can see that the condition expressed by Eq. (\ref{condition})  can
be restated, for modes $ k \leq \omega_{p}$, as  $|
\vec{E}_{\vec{k},\omega}|^2 < n_q T$ (where we essentially took the
square modulus of Eq. (\ref{cond})). This last inequality expresses
the fact that the energy  density associated with the gauge field
fluctuations  should always be smaller than the critical energy
density  stored in radiation. The linear treatment of the Vlasov
equation is certainly  accurate provided the typical modes of the the
field are  smaller than the plasma frequency and provided  the energy
density in electric and magnetic fields is smaller  than $T^4$, i.e.
the energy density stored in the radiation  background.

\end{appendix}

\end{document}